\documentclass[aps,article,twocolumn,superscriptaddress,english]{revtex4}
\usepackage{graphicx}
\usepackage{amsmath}
\usepackage{epsfig}
\usepackage{color}
\usepackage{xcolor}
\usepackage{ulem}
\usepackage{lineno}

\usepackage{amssymb}
\usepackage{graphicx}
\usepackage{amsmath}
\usepackage{amsfonts}
\usepackage[T1]{fontenc}
\usepackage[latin9]{inputenc}
\usepackage{array}
\usepackage{multirow}
\usepackage{color}
\usepackage{esint}
\usepackage{bm}
\usepackage{bbm}
\usepackage{hyperref}
\usepackage{titlesec}
\usepackage{soul}
\usepackage{mathtools}
\usepackage{amsmath}

\newcommand{\ket}[1]{\ensuremath{|{#1}\rangle}}

\newcommand{\red}[1]{\textcolor{black}{#1}}
\newcommand{\blue}[1]{\textcolor{black}{#1}}

\begin{document}

\title{Observing a topological phase transition with deep neural networks from experimental images of ultracold atoms }

\author{Entong Zhao}
\thanks{Theres authors contributed equally.}
\affiliation{Department of Physics, The Hong Kong University of Science and Technology,\\ Clear Water Bay, Kowloon, Hong Kong, China}

\author{Ting Hin Mak}
\thanks{Theres authors contributed equally.}
\affiliation{Department of Physics, The Hong Kong University of Science and Technology,\\ Clear Water Bay, Kowloon, Hong Kong, China}

\author{Chengdong He}
\affiliation{Department of Physics, The Hong Kong University of Science and Technology,\\ Clear Water Bay, Kowloon, Hong Kong, China}

\author{Zejian Ren}
\affiliation{Department of Physics, The Hong Kong University of Science and Technology,\\ Clear Water Bay, Kowloon, Hong Kong, China}

\author{Ka Kwan Pak}
\affiliation{Department of Physics, The Hong Kong University of Science and Technology,\\ Clear Water Bay, Kowloon, Hong Kong, China}

\author{Yu-Jun Liu}
\affiliation{Department of Physics, The Hong Kong University of Science and Technology,\\ Clear Water Bay, Kowloon, Hong Kong, China}

\author{Gyu-Boong Jo}
\email{gbjo@ust.hk}
\affiliation{Department of Physics, The Hong Kong University of Science and Technology,\\ Clear Water Bay, Kowloon, Hong Kong, China}
\affiliation{IAS Center for Quantum Technologies, Clear Water Bay, Kowloon, Hong Kong, China}
\affiliation{Microelectronics Thrust, The Hong Kong University of Science and Technology (Guangzhou), Guangzhou, China}

\begin{abstract}
Although classifying topological quantum phases have attracted great interests, the absence of local order parameter generically makes it challenging to detect a topological phase transition from experimental data. Recent advances in machine learning algorithms enable physicists to analyze experimental data with unprecedented high sensitivities, and identify quantum phases even in the presence of unavoidable noises. Here, we report a successful identification of topological phase transitions using a deep convolutional neural network trained with low signal-to-noise-ratio (SNR) experimental data obtained in a symmetry-protected topological system of spin-orbit-coupled  fermions. We apply the trained network to unseen data to map out a whole phase diagram, which \blue{predicts the positions of the two topological phase transitions} that are consistent with the results obtained by using the conventional method on higher SNR data.  By visualizing the filters and post-convolutional results of the convolutional layer,  \blue{we further find that the CNN uses the same information to make the classification in the system as the conventional analysis, namely spin imbalance, but with an advantage concerning SNR.} Our work highlights the potential of machine learning techniques to be used in various quantum systems.\end{abstract}

\maketitle

\section{Introduction}
Since the discovery of quantum Hall effect in a 2D electron gas~\cite{klitzing:1980nm},  topological quantum phases, which are usually characterized by nonlocal invariants, have played a pivotal role in quantum matter research. In contrast to the solid-state materials, ultracold atoms provide a clean and highly-controllable experimental platform to investigate topological quantum systems~\cite{Bloch:2008gl, Goldman:2016fa}.  To date, many paradigmatic topological models have been realized and explored in experiments with ultracold atoms using artificial gauge fields~\cite{Aidelsburger.2018,Zhang.2018}, such as the realization of  the 1D Su-Schrieffer-Heeger model~\cite{Atala:2014gf}, the 1D symmetry-protected topological (SPT) phase~\cite{Song:2017uf},  the 2D Chern insulator~\cite{Wu:2016rt} and nodal~\cite{Song.2019} or Weyl~\cite{Wang.2020avm} semimetals in 3D. \blue {While some topological phases have already been successfully characterized in recent works ~\cite{Wu:2016rt,Flaschner.2016,Song:2017uf,Song.2019}, uncovering a proper observable is generically subtle and highly model-dependent due to the absence of the local order parameter, which makes detecting topological phases from the experimental data obtained from ultracold atoms remains challenging. }
	
%While the realization of such models complement the study of condensed matter physics, detecting topological phases from the experimental data obtained from ultracold atoms is still challenging although some topological phases are successfully characterized in recent works~\cite{Wu:2016rt,Flaschner.2016,Song:2017uf,Song.2019}. This challenge stems from the fact that uncovering a proper observable is subtle and highly model-dependent due to the absence of the local order parameter.

Recently, machine learning techniques, such as deep convolutional neural network, have become an indispensable tool in many technological areas and also proven its power in quantum many-body physics~\cite{Wang.2016,Zhang.20191q, Carleo2017, Bohrdt:2019fu, Zhao.2020}. \blue {For example, both supervised learning and unsupervised learning have been used to determine topological phases~\cite{Zhang.2018e8} and further map out the phase diagram \cite{Rem:2019hx, kaming2021unsupervised} due to their ability to classify or identify massive data sets such as experimental images}. Here we implement machine learning techniques such as deep convolutional neural networks (CNN) to classify single-shot absorption images with different SPT phases which are characterized by the $\mathbb{Z}_2$ invariant~\cite{Song:2017uf}. We train the neural network on experimental images far from the phase transition points, and then use the trained neural network to identify the phase transition point. \blue{An exemplary analysis with a conventional method demonstrates that the classification of single pair of images is difficult for conventional techniques and sufficiently large ensemble average is required due to the low SNR in single pair of images.} By visualizing the filters and activation region of the trained neural network, we further interpret that spin imbalance information plays an important role when the neural network makes the classifications. Our work demonstrates the ability of machine learning techniques in processing experimental images which may provide a deeper understanding of the emergent topological quantum matter.

\begin{figure}%[htbp] 
	\centering
	\includegraphics[width=1\linewidth]{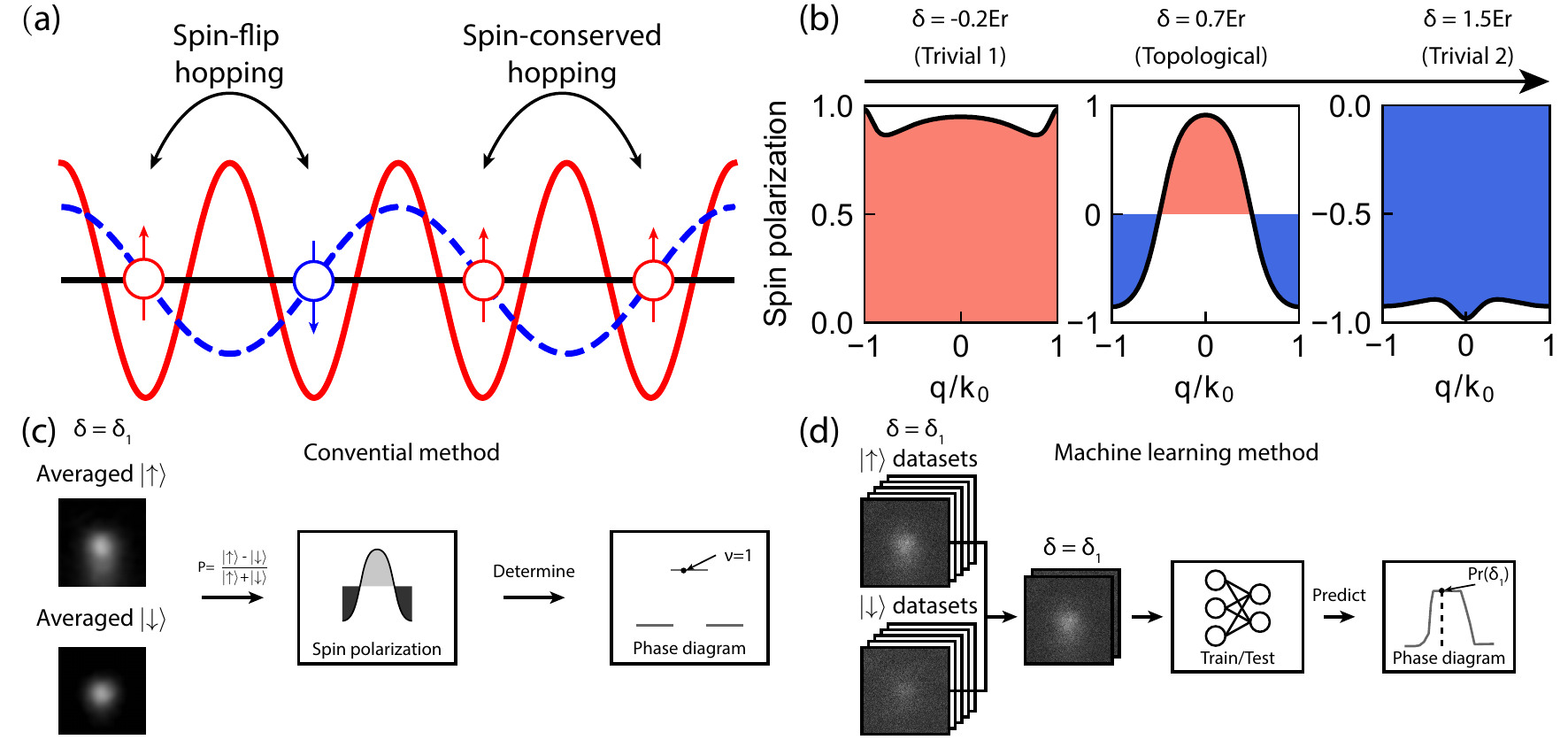}
	\caption{{\bf Topological optical lattices and data analysis with machine learning} (a) In the 1D optical Raman lattice model,  a Raman potential (blue dashed line) induces spin-flip hopping. whereas  a lattice potential (red solid line) contributes to spin-conserved hopping. (b) A typical {\color{black}spin polarization} of the ground band is shown for topological and trivial regimes.  (c) A conventional analysis requires a set of images taken for each spin, resulting in a statistically averaged density distribution.  A {\color{black}spin polarization}  is then calculated from those images, and a topological invariant is determined. (d) The neural network is trained with experimental images far from the phase transition points without obtaining averaged {\color{black}spin polarization}  images.  The phase transition point is then identified by the trained neural network with a single shot.}
	\label{Fig:0}
\end{figure}

\section{Experimental preparation}

The topological phase transition to be analyzed by the neural network is experimentally realized with ultracold fermions in {\color{black} an optical lattice with spin-orbit couplings as described in Fig.~\ref{Fig:0}a~\cite{Liu:2013wm,Zhang.2018olh} (see Supplementary materials for more details)}.  We begin with pre-cooled $^{173}$Yb atoms in the intercombination magneto-optical trap followed by the evaporative cooling in a crossed optical dipole trap. After the final stage of the optical evaporative cooling, we achieve a two-component Fermi gas of $N=1.0\times 10^4$ atoms in $\ket{\uparrow}=\ket{m_F=\frac{5}{2}}$  and $\ket{\downarrow}=\ket{m_F=\frac{3}{2}}$ at $T/T_F\simeq 0.4$ where $T_F$ is the Fermi temperature~\cite{Song.2016be,Song:2017uf}. The atoms are then loaded into a 1D optical  lattice dressed by a Raman coupling potential within 10 ms. {\color{black}A spin-dependent optical ac Stark shift is added to energetically lift up all other hyperfine levels except two relevant levels realizing  
an effective spin-1/2 subspace. This Stark shift beam is applied 5ms before the optical lattice potential and the Raman coupling beam are adiabatically ramped up.} After holding the atoms in lattice for 2ms, all the laser beams are suddenly switched off and 8ms spin-resolved time-of flight (TOF) images are taken after a spin sensitive blast sequence {\color{black}(see Supplementary materials for more details)}. Therefore, for each data we obtain three raw images which include an image $\mathcal{I}_{\downarrow,o}$ with $\ket{\uparrow}$ atoms removed, an image $\mathcal{I}_{\uparrow,o}$ with $\ket{\downarrow}$ atoms removed and a background image $\mathcal{I}_o$ with both $\ket{\uparrow}$ and $\ket{\downarrow}$ atoms removed. Finally, the distribution of spin $\ket{\uparrow}$ and  $\ket{\downarrow}$ atoms can be extracted from $\mathcal{I}_{\uparrow,o}-\mathcal{I}_o$ and $\mathcal{I}_{\downarrow,o}-\mathcal{I}_o$, respectively(see Fig.~\ref{Fig:0}c,d).

The realized 1D optical Raman lattice \blue{has} a Hamiltonian of 
\begin{equation*}
 \begin{aligned}
H &=[\frac{p_x^2}{2m}+\frac{V_{\uparrow}(x)+V_{\downarrow}(x)}{2}]\otimes\mathbf{\hat{1}}+[\frac{\delta}{2}+\frac{V_{\uparrow}(x)-V_{\downarrow}(x)}{2}]\mathbf{\sigma_z} \\ 
&+\mathcal{M}(x)\mathbf{\sigma_x}
\end{aligned}
\end{equation*} 

\noindent where $\frac{p_x^2}{2m}$ is the kinetic energy along x direction, $V_{\uparrow, \downarrow}(x)=V_{0\uparrow,0\downarrow}\cos^2{(k_0x)}$ are the optical lattice potentials for spin $\ket{\uparrow,\downarrow}$ states with $V_{0\uparrow,0\downarrow}$ denote the lattice depths and $k_0=\pi/a$, $\mathcal{M}(x)=M_0\cos{(k_0x)}$ is the periodic Raman coupling potential with amplitude $M_0$, $\delta$ is the two photon detuning and $\mathbf{\hat{1}}, \mathbf{\sigma_{x,y,z}}$ denote the unit matrix and Pauli matrices. In the Hamiltonian, the lattice potential $V_{\uparrow, \downarrow}(x)$ induces the nearest-neighbor hopping which conserves the spin, while the Raman coupling term $\mathcal{M}(x)$ contributes to the hopping that flips the spin. The effect of these hoppings leads to nontrivial topological phases protected by symmetries~\cite{Song:2017uf}. According to the {\color{black}Altland-Zirnbauer classification~\cite{Altland:1997cc}} and further calculation, the Hamiltonian satisfies a nonlocal chiral symmetry and a magnetic group symmetry that is defined as the product of time-reversal and mirror symmetries, and the topological invariant of this system is characterized by an integer winding number $\mathcal{N}_{1D}=\nu\cdot\rm{sgn}[P(\Gamma)]$, where $\nu$ is the $\mathbb{Z}_2$ invariant satisfying $(-1)^{\nu}=\prod_j\rm{sgn}[P(\Lambda_j)]$, and $P(\Lambda_j)=\pm1$ is the spin polarization at two symmetric momenta $\{\Lambda_j\}=\{\Gamma(q_x=0),M(q_x=\pi/a)\}$  in the first Brillouin zone (FBZ)~\cite{Liu:2013wm,Altland:1997cc,Song:2017uf}. By varying the two-photon detuning $\delta$, we can realize the topologically nontrivial or trivial phases corresponding to $\nu=1$ or $\nu=0$ {\color{black}as shown in Fig.~\ref{Fig:0}b}.

\begin{figure}[htbp] 
	\centering
	\includegraphics[width=\linewidth]{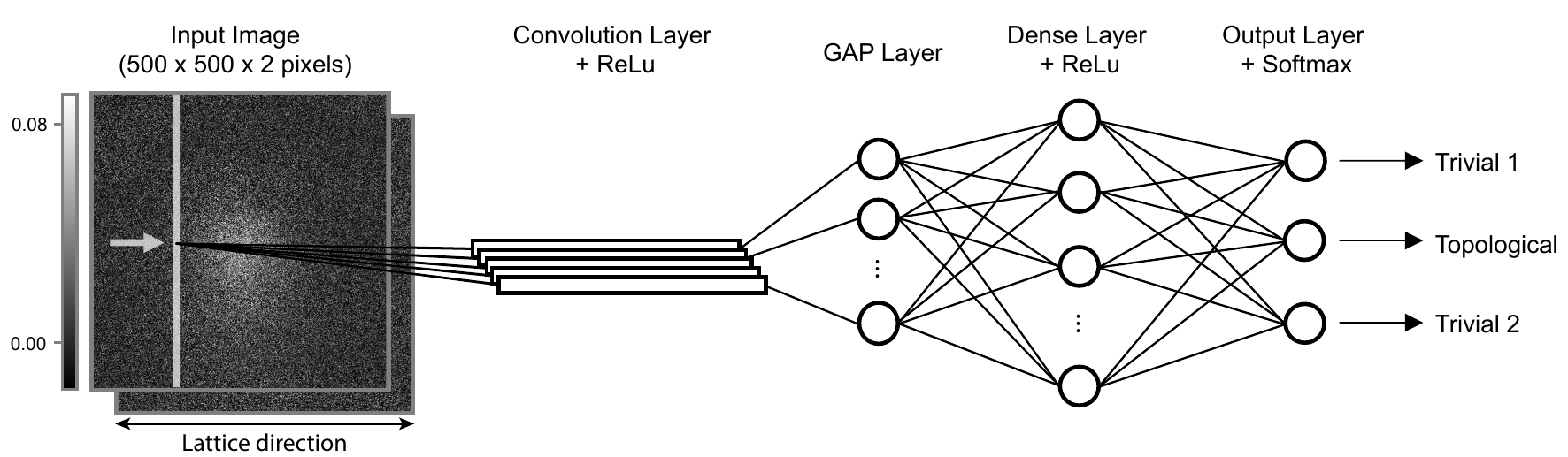}
	\caption{Absorption images for $\ket{\uparrow}$ and $\ket{\downarrow}$ atoms after 8ms expansion form the two channels of the input for the CNN with a convolutional layer, a GAP (Global Average Pooling) layer and three hidden dense layers. The gray arrow represents the sliding of the vertical 500 $\times$ 1 filters across the input image. The softmax activation function outputs the probability of an image of belonging to one of the classes, where +1, 0, -1 means the Trivial 1, Topological, Trivial 2 class respectively. The CNN is trained and validated on two sets of labeled images, which enables it to classify unseen images with high confidence.}
	\label{Fig:1}
\end{figure}

\section{Train neural network to detect the phase transition}
\begin{figure*}[htbp] 
	\centering \includegraphics[width=0.8\linewidth]{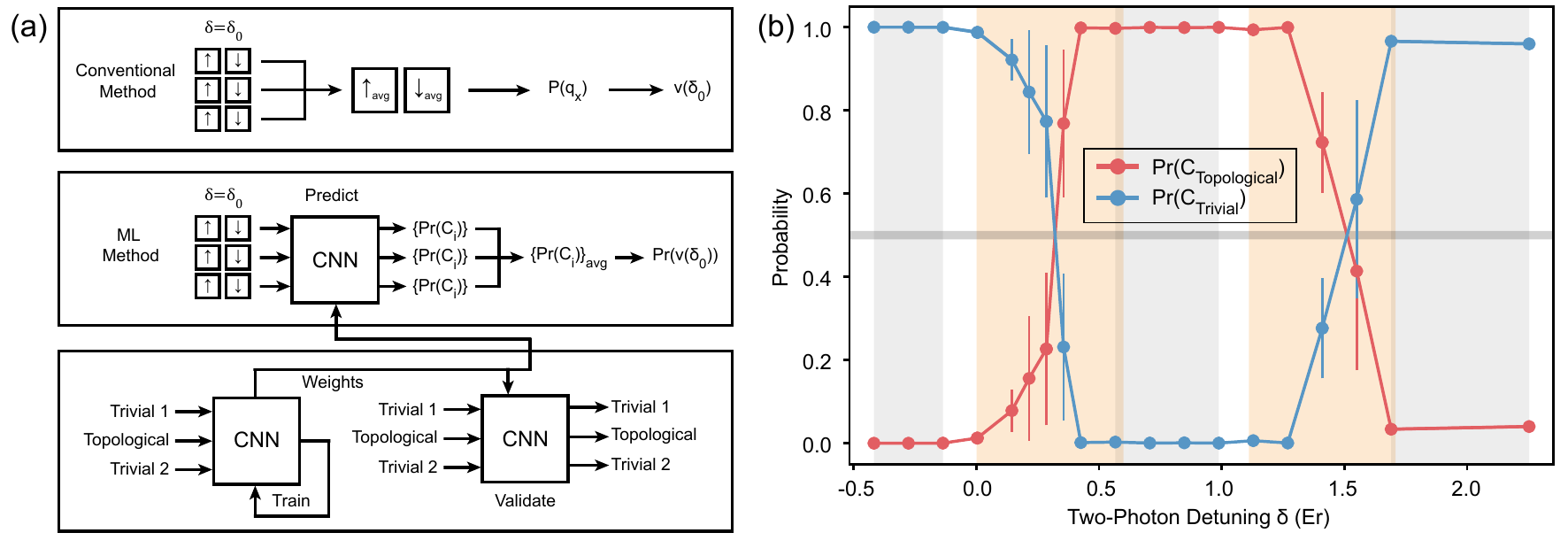}
	\caption{(a) A schematic diagram of obtaining a point of the phase diagram (the topological phase $\nu$ with respect to a detuning $\delta_0$) via the conventional method and ML method separately. In the conventional method, experimental images of same detuning are first averaged to calculate the {\color{black}spin polarization}  $P(q_x)$, the phase $\nu(\delta_0)$ is then obtained from the behavior of $P(q_x)$ in the FBZ. In the ML method, experimental spin up and down image pairs from the test dataset are fed into the trained CNN to output a corresponding predicted phase probabilities. The phase probabilities of image pairs with same detuning are finally averaged to give the probability of the detuning $Pr(\nu(\delta_0))$. Prior to making predictions on test dataset images, the CNN is trained and validated with two other individual datasets. The training, validation, and test dataset, with 140, 31 and 95 datapoints respectively. The training and validation dataset contains only datapoints from deep topological or deep trivial regimes only (gray shadows in Fig. 2b), while the test dataset contains all detuning values (5 datapoints per detuning value). (b) Predicted phase diagram by the trained CNN on the test dataset. The orange shadow is the transition regime from theoretical prediction at finite temperature with uncertainty of the two-photon detuning being $0.3E_r$. By identifying the 50\% probability point as the phase transition, both predicted transition points are consistent with the experimental calculations.  Error bars are shown in standard error.}
	\label{Fig:2}
\end{figure*} 

We take the spin $\ket{\uparrow}$ and $\ket{\downarrow}$ data for varying two-photon detuning. The $\ket{\uparrow}$ and $\ket{\downarrow}$ of each experimental run constitutes a sample for the model. The label of the data can be determined by normalized spin polarization $P(q_x)=(n_\uparrow(q_x)-n_\downarrow(q_x))/(n_\uparrow(q_x)+n_\downarrow(q_x))$ averaged over all the data with same two-photon detuning $\delta$. When $P(q_x)>0\quad\forall q_x\in \text{FBZ}$ or $P(q_x)<0 \quad\forall q_x\in \text{FBZ}$, the phase is topologically trivial with the $\mathbb{Z}_2$-invariant $\nu=0$ and we label the data to Trivial 1 or Trivial 2 respectively. For the case that $P(q_x)$ is partially spin-up dominated, the phase is topologically nontrivial with the $\mathbb{Z}_2$-invariant $\nu=1$ and we label the data to Topological. For each detuning value, five data are randomly chosen to form the test dataset and the remaining data with the detuning value far from the transition regimes are randomly assigned to a train dataset and a validation dataset with an approximate 4:1 ratio. The resultant train, validation and test dataset contains 140, 31 and 95 datapoints respectively.

We then establish a CNN architecture that contains a convolutional layer, a Global Average Pooling (GAP) layer and three hidden fully connected layers {\color{black}as shown in Fig.~\ref{Fig:1}}. We use 32 1D filters in the convolutional layer oriented perpendicular to the direction of the lattice direction. The choice for the orientation of filters is based on the physical intuition that the system is one-dimensional, thus the filter correlates data in the vertical direction and \blue{passes} a one-dimensional information along  the lattice (horizontal) direction to the rest of the CNN, similar to the integration procedure before analyzing in the conventional method. During training process, the train dataset is fed into the neural network which outputs a prediction probability $Pr(C_i)$ for each class $\{C_i\} = \{ \text{Trivial 1, Topological, Trivial 2} \}$ when {\color{black}a single pair of images} is loaded. The parameters of the CNN are tuned iteratively to minimize the sparse categorical cross entropy loss function using the Adam optimizer \cite{Kingma:2014}. \blue{After every epoch of training, the model is evaluated with the validation dataset, and the training is completed if the validation accuracy does not further improve. The final train and validation accuracy are both 100$\%$, and the total number of epochs used is around 120.}

\begin{figure}[htbp] 
	\centering \includegraphics[width=\linewidth]{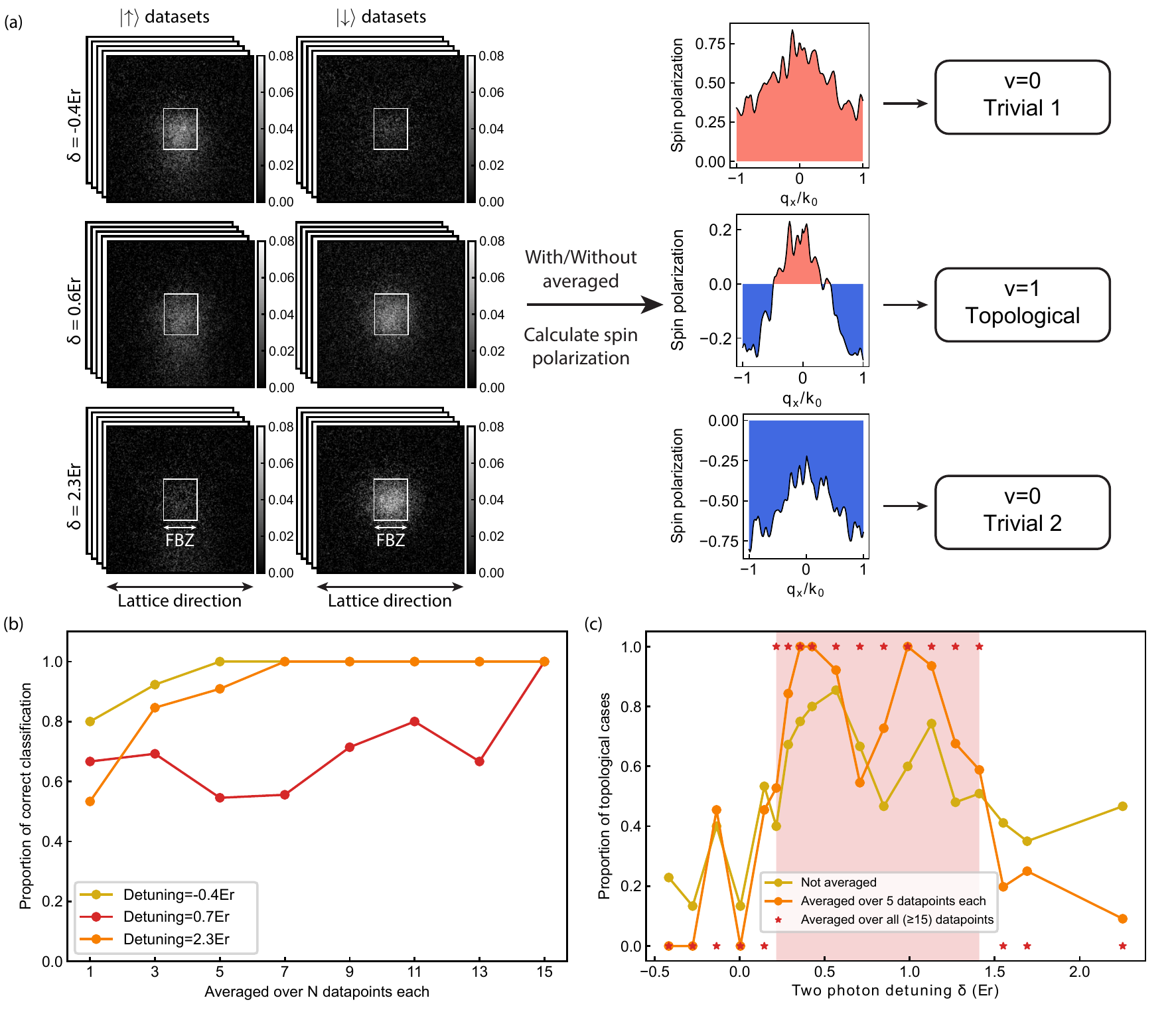}
	\caption{\textbf{An exemplary analysis using conventional method.} (a) Procedure to determine the trivial or topological band using conventional analysis is shown for three pairs of images in three different classes (Trivial 1, Topological, Trivial 2). Both of the density distributions and the normalized spin polarization data shown here are without averaged. The white rectangles in the single shot density distributions are the regions of interest for conventional analysis, with the horizontal axis representing the first Brillouin zone (FBZ). (b) The proportions of correct classification at three detuning values far from the transition regime with different numbers of ensemble average are used to evaluate the performance of conventional method. (c) Phase diagrams are obtained from data without averaged, ensemble averaged \red{over 5 datapoints each} and ensemble averaged over all data (larger than 15 \blue{datapoints} per detuning value). For the ensemble averaged over all data, the proportion is either 0 or 1 and the predicted topological region is shown in red shaded region.}
	\label{Fig:3_advantage}
\end{figure}  

To identify the transition regime, we let the trained CNN predict the reserved test dataset. Specifically, we sum up the predicted probabilities for Trivial 1 and Trivial 2 classes to obtain the probability for topologically trivial case (Fig. 2a), since they have the same $\mathbb{Z}_2$ invariant $\nu=0$, and the predicted probability for Topological class corresponds to $\mathbb{Z}_2$ invariant $\nu=1$ {\color{black}as summarized in Fig.\ref{Fig:2}a}. The output probability for $\nu=1(0)$ is then plotted as a function of two-photon detuning $\delta$ (Fig.\ref{Fig:2}b), providing two phase transition regime around $\delta=0.32\text{Er}$ and $\delta=1.51\text{Er}$. The prediction is in quantitative agreement with our previous conventional analysis \cite{Song:2017uf}, which is extracted from another dataset with shorter TOF and higher signal-to-noise ratio. (shown by the pink shadow in Fig.\ref{Fig:2}b)

{\color{black}To highlight the advantage gained via the neural network approach, we show an exemplary analysis using the conventional method in Fig~\ref{Fig:3_advantage}. A spin polarization of the ground band is reconstructed from spin-up and spin-down averaged images (Fig.~\ref{Fig:3_advantage}a) leading to the measurement of topological invariant. However, sufficiently large ensemble average is required to correctly classify the topological states as shown in Fig.~\ref{Fig:3_advantage}b. \red{For detuning $\delta=0.7$Er, the classification from the conventional method hardly converges before an averaging of 15, which indicates even more images are needed for an unambiguous classification.} Notably, it is almost impossible to probe the topological phase transition using the conventional analysis with a single pair of images without ensemble averaging (Fig.~\ref{Fig:3_advantage}c). }

\begin{figure}[htbp] 
	\centering \includegraphics[width=\linewidth]{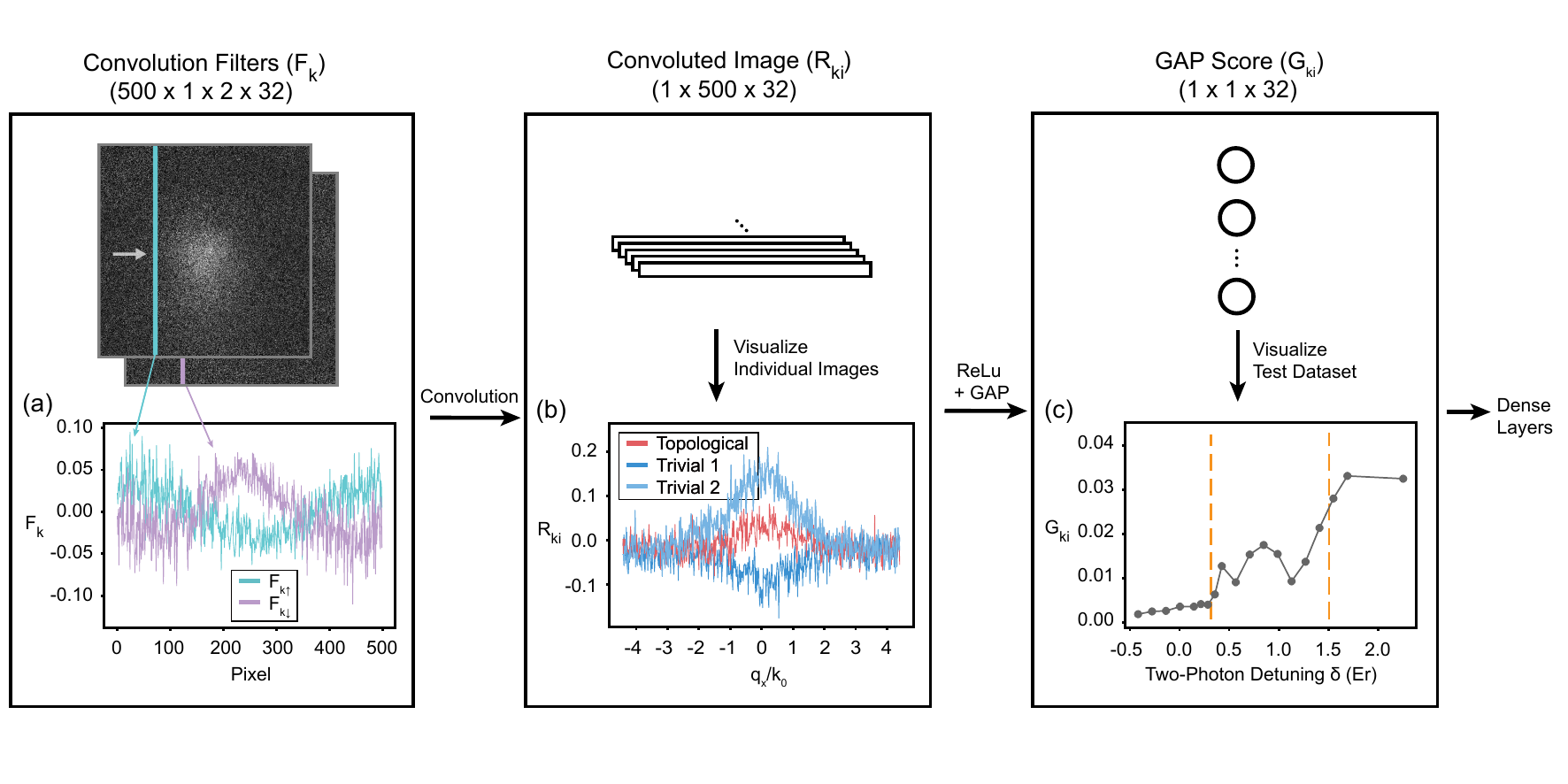}
	\caption{\textbf{Interpretation of the neural network.} (a) Visualization of the weight of a  filter of the trained CNN. Approximately 75\% of the filters are similar to this filter up to switching the two channels. {\color{black} Here, we show the result for the filter $k$=11.} (b) Illustration of the effect of the filter by visualizing the post-convolution results $R_{ki}$. An image of each class is processed (convoluted) by the filter shown in (a). A comparison with the experimental calculation shows that the effect of this filter is to perform an approximation of  $\pm(n_\uparrow - n_\downarrow)$ of the input image.  (c) Illustration of the effect of the GAP layer by visualizing the post-GAP layer results of all datapoints in the test dataset, again with the filter in (a). The ReLu operation truncates the negative part of the graphs, while the GAP layer finds the normalized area to reduce the dimension for classification. When plotted against detuning value using the test dataset, three different plateaus are formed with two boundaries locating near the phase transition regime, such that the phase classification can now be easily done by identifying two critical values for $G_{ki}$. The orange dotted lines are the predicted transition points by the model (the intersecting points in Fig. 3b).}
	\label{Fig:3}
\end{figure}

\section{Interpret the neural network}

Whereas we have focused on the machine learned analysis of topological phases transition, it will be interesting to investigate how the neural network extracts physical observables relevant to the topological property of the system. This understanding may \blue{to some extent exclude the possibility that the probability curve \red{in Fig.\ref{Fig:2}b} is a simple interpolation, and also} allow to generalize the application of machine learning analysis to various many-body quantum systems~\cite{Zhao.2020, miles2021correlator, khatami2020visualizing, ponte2017kernel, greitemann2019probing, dawid2020phase, zhang2020interpreting, wetzel2020discovering}. In our work, the machine learning analysis of topological quantum phases can guide us to extract right features from experimental images. 

Here, we investigate how the neural network performs the classification in our system. To this end, we start by visualizing the filters $F_k$ $(k = 1,...,32)$ of the optimal CNN. Each filter consists of 2 channels $F_{k\uparrow}$ and $F_{k\downarrow}$ which act on spin up and spin down image respectively and serve as integration operations on the vertical dimension of the corresponding image. Looking at the filters, we find that approximately 75\% of the filters are trained to have an opposite weight for spin $\ket{\uparrow}$ and $\ket{\downarrow}$  channels (One example with $k=11$ is shown in Fig. \ref{Fig:3}a; {\color{black}all filters being used in this procedure are availabe in Fig.~S2 of Supplementary material}). This type of subtraction operation indicates that these filters attempt to learn the spin imbalance information from the spin $\ket{\uparrow}$ and $\ket{\downarrow}$ data, which is one of key observables in topological bands for ultracold atoms~\cite{Song:2017uf}. Besides, the channels also display opposite sign of weights between the central and background regions, which shows that the filter learns to distinguish the central region from the background noises and suppresses the noises. The weights in the central region have large magnitudes, which emphasizes the central region of the image during the integration. Analogously, in the conventional analysis, we first preprocess the image by cropping the central 141 pixels along the vertical dimension before calculating the spin imbalance.

We further pass three images from three different classes $I_i$ through the convolutional layer of the neural network to calculate the post-convolution results $R_{ki}$ for this filter $F_k$, which is also known as the feature maps. (Fig. 4b),
\begin{equation*}
 \begin{aligned}	
R_{ki}(x') &=I_i(x')\cdot F_k+B_k \\ &= \sum_{y} [F_{k\uparrow}(y) I_{i\uparrow}(x',y) + F_{k\downarrow}(y) I_{i\downarrow}(x',y)] + B_k
 \end{aligned}
\end{equation*}

where $F_{k\uparrow}$ and $F_{k\downarrow}$ denotes the weights of filters for spin $\ket{\uparrow}$ and $\ket{\downarrow}$ channels respectively, $B_k$ is the bias corresponding to the filter $F_k$. This type of filter outputs distinguishable feature maps for three different classes, with magnitude Trivial 2 $>$ Topological $>$ Trivial 1, which looks like conventionally-calculated spin polarization within the FBZ. These results suggest that the behavior of the convolutional layer is analogous to the conventional analysis, where the spin polarization is first extracted from the spin $\ket{\uparrow}$ and $\ket{\downarrow}$ atoms. 

Finally, we record the averaged GAP score $G_{ki}= \frac{1}{500} \text{ReLu}(\sum_{x'} R_{ki}(x'))$ when all the data in test dataset forward-propagate the first stage of the network. The resultant $G_{ki}$ is then plotted as a function of two photon detuning $\delta$ as shown in Fig.\ref{Fig:3}c. Since $G_{ki}$ can be understood as the average value of the negative-truncated area in $R_{ki}$, it takes a value of nearly zero, relatively high positive, and intermediate when the input image belongs to Trivial 1, Trivial 2, and Topological class respectively. This forms three different plateaus with two boundaries locating near the phase transition regime. Thus, the GAP score outputted by each filter can be understood as a single-filter version phase classifier and the dense layers further process, compare and summarize the independent viewpoints from all filters and give a final probability with \blue{an} improved validity to reduce the appearance of outliers. (Visualization of other filters \blue{is} listed in the supplemental document.)

\section{Conclusion}
In summary, we have demonstrated the power of machine learning techniques in processing massive dataset by applying a deep CNN to the experimental images with different SPT phases characterized by the $\mathbb{Z}_2$ invariant. The trained neural network identifies two phase transition regimes, which is in good agreement with the previous result processed with conventional analysis. Different from the previous 6~ms TOF dataset in \cite{Song:2017uf}, the current dataset has longer TOF, which potentially leads to lower signal-to-noise ratio. Except for the training dataset and their labels, only limited prior knowledge is required when we establish the network architecture and pre-process the dataset, which provides better generalization capability comparing with conventional analysis. We further visualize the filters and convoluted result of the trained neural network and find that the spin imbalance information plays a significant role when the neural network \blue{makes} the classification\blue{, which implies that the probability curve represents the phase transition between two distinct phases instead of simple interpolation}. {\color{black} This suggests that the neural network analysis not only processes previously unknown information~\cite{Zhao.2020} but also \blue{performs} a conventional analysis in a noise-resilient manner. It will be interesting to systematically investigate how the neural network can be resilient to systematic noises (e.g. shot-to-shot shift of the atomic cloud or photon shot noises) than conventional analysis. }\blue{Another possible future direction is unsupervised machine learning algorithms which can identify the phase transition with unlabeled data and be applied to quantum systems with unknown topological phase diagrams or hidden orders\cite{kaming2021unsupervised}.}

Our work shows machine learning techniques, such as CNNs, provide a \blue{useful} and convenient tool to analyze the \blue{experimental data with limited SNR} obtained from ultracold atom experiments. In particular, this technique would be particularly useful for analyzing spin polarization in topological systems~\cite{Flaschner.2016,Song.2019,Ren.2022}. Especially, this approach may open an interesting possibility of extracting topological invariants from the high-dimensional topological system~\cite{Song.2019}.

\newpage

\paragraph*{\bf Acknowledgement} We acknowledge support from the RGC through 16304918, 16308118, 16306119, 16302420, 16302821, 16306321 C6005-17G, C6009-20G and RFS2122-6S04. This works is also supported by the Guangdong-Hong Kong Joint Laboratory and the Hari Harilela foundation.

\paragraph*{Disclosures} The authors declare no conflicts of interest.

\paragraph*{Supplementary information} See Supplement 1 for supporting content.

\paragraph*{Data availability} Data underlying the results presented in this paper are not publicly available at this time but may be obtained from the authors upon reasonable request.

\end{document}